# Three-dimensional stochastic retinal models for synthetic eye-models

Nadine P. van Dam[1,2,3], Luc van Vught[1], Jos J. Rozema[4,5], Steven J. M. Habraken[3,6] , Jan-Willem M. Beenakker[1,2,3,6]

[1]Department of Ophthalmology, Leiden University Medical Center, Leiden, The Netherlands
[2]Department of Radiology, Leiden University Medical Center, Leiden, The Netherlands
[3]Department of Radiation Oncology, Leiden University Medical Center, Leiden, The Netherlands
[4]Department of Ophthalmology, Antwerp University Hospital, Edegem, Belgium
[5]Visual Optics Laboratory Antwerp, Faculty of Medicine and Health Sciences, University of Antwerp, Wilrijk, Belgium
[6]HollandPTC, Delft, the Netherlands

**Corresponding Author:**
Nadine P. van Dam
Department of Ophthalmology
Leiden University Medical Center
Albinusdreef 2
Leiden, 2333 ZA, The Netherlands
n.p.van_dam@lumc.nl

**Funding:** Dutch Research Council (NWO), Grand Number: OTP-CAPTOR, 21360.

**Commercial Relationship Disclosure:** Nadine P. van Dam, Trident S.r.l., and MedCom GmbH; Luc van Vught, Current employee of  Cassini Technologies B.V.; Jos J. Rozema, Consultant for Azalea Vision and Morrow Vision; Steven J. M. Habraken, Trident S.r.l., and MedCom GmbH; Jan-Willem M. Beenakker, Philips Healthcare, Trident S.r.l., and MedCom GmbH.

# Abstract

**Purpose:** Current eye models often assume a fixed spherical retinal curvature of 12 mm, which does not reflect the anatomical variability observed in vivo. This study proposes an extension incorporating the stochastic relationship between axial length and three-dimensional (3D) retinal curvature, enabling more realistic 3D eye models.

**Methods:** Axial length and 3D retinal curvature data from 94 subjects, split into five subsets for cross-validation, were used to construct two probabilistic models: a Single Gaussian Model (SGM) and two-component Multi Gaussian Mixture Model (MGMM). Both were converted into conditional forms to estimate retinal curvature for a specific axial length, and evaluated against measured data distributions.

**Results:** All models produced radii of curvature comparable to measured data (median difference <0.2 mm). Non-conditionally, the MGMM described the distribution slightly better than SGM, with higher density overlap (>94% vs. >86%), but similar variance (worst f-test (p-value) 1.05 (0.71) vs. 1.05 (0.73)). Conditioned on axial length, both models performed similarly with overlap exceeding 93% and the worst f-test 1.02 (0.84). Given its simplicity and robustness, the conditional SGM was selected, achieving similar validation results: median difference (<0.10 mm), overlap (>92%), and a non-significant worst f-test (1.02 (0.85)).

**Conclusions:** The conditional SGM enables accurate generation of anatomically realistic 3D retinal curvatures based on axial length, supporting improved stochastic eye modeling for visual optics and population-based simulations of healthy eyes.

**Translational relevance:** Replacing the fixed 12 mm retina assumption with a population-based stochastic model improves anatomical accuracy of eye models used in optical design and treatment planning.

## Introduction

An accurate model of the ocular geometry is essential for studying the optical properties of the eye and developing effective therapeutic strategies. Over the years, several schematic eye models have been proposed, including the Gullstrand [1], Le Grand [2], Emsley [3], and Navarro [4] models [5], which rely on a single set of parameters to describe ocular surfaces. Each of these models was designed to accurately capture a specific property of the eye, such as higher-order aberrations or peripheral refraction. These models have served as a foundation for understanding vision complaints such as negative dysphotopsia and for guiding improved treatment strategies, including intra-ocular lenses that maintain peripheral vision quality [6-11].

In contrast to these schematic models, real eyes exhibit a wide variety of anatomical variations [12-15]. The statistical relationships between key biometric parameters, such as axial length, corneal curvature, and anterior chamber depth, have been extensively documented [16-20] and form the basis of nomograms that guide clinical decision- making, such as selecting an intraocular lens for cataract surgery. Rozema et al. introduced SyntEyes [21,22], a stochastic eye model that unlike conventional eye models, considers the statistical distribution of biometric parameters and their correlations. Instead of specifying a single eye model, a stochastic eye model generates a virtual cohort of eyes that realistically capture the variability in ocular biometry within the normal population. These generated eyes have been used in vision science to simulate the correction of refractive errors in healthy eyes [23] or eye tracking [24] with different biometric parameters, without the need for real clinical data. Additionally, SyntEyes have also been designed for specific clinical conditions such as keratoconus [25], and were used for similar corrective simulations [26].

Although the SyntEyes model captures most ocular biometry parameters relevant for central vision, they assume a spherical retina with a fixed 12 mm radius of curvature, similar to other eye models [27,28]. However, several studies have shown that the retinal surface is more accurately represented by an ellipsoidal shape and that its radii differ significantly between individuals and from the generic value of 12 mm, with standard deviations up to 0.7 mm [9,29-32]. These variations in retinal geometry influence clinically relevant functional characteristics, including peripheral refraction [32], prediction of myopia progression [33], and the quantification of fundus imaging [34]. Consequently, the applicability of SyntEyes for assessment beyond central vision is limited.

It is hypothesized that the Multivariate Gaussian Mixture Model (MGMM), which underpins the SyntEyes framework, can also be applied to accurately describe the statistical relationship between axial length and the retinal radii, as measured by high-resolution Magnetic Resonance Imaging (MRI). Leveraging this approach would enable the extension of SyntEyes to incorporate three-dimensional retinal shape information, providing a more anatomically accurate representation of the posterior segment of the eye.

# Methods

*Participants*

In this study, 94 individuals with no relevant ophthalmic history (e.g. no pathology or condition), except for cataract surgery in a subset of participants were retrospectively included. Ocular biometry (Lenstar LS900; Haag-Streit AG, Koeniz Switzerland), with i.e. the axial length defined as the distance from the cornea to inner limiting membrane, and high-resolution MRI scans were prospectively acquired in the context of other clinical studies, each of which adhered to the Declaration of Helsinki and received approval by the Medical Ethics Committee Leiden The Hague Delft [6,9,10,29]. All the participants provided written informed consent before their inclusion into the studies.

The retinal curvature parameters were derived from high-resolution MRI, as previously described [9,29]. All MRIs were acquired with a Philips Achieva 7T whole-body magnet (Philips Healthcare, Best, The Netherlands) equipped with an in-house developed eye receive coil [9,29,35]. Retinal segmentation was carried out automatically using subdivision surfaces in MeVisLab (MeVis Medical Solutions AG, Bremen, Germany) [9,29]. Subsequently, a three-dimensional ellipsoid was fitted to the posterior 240 degrees of the retinal contour, following the method proposed by van Vught et al. [29], Figure 1.

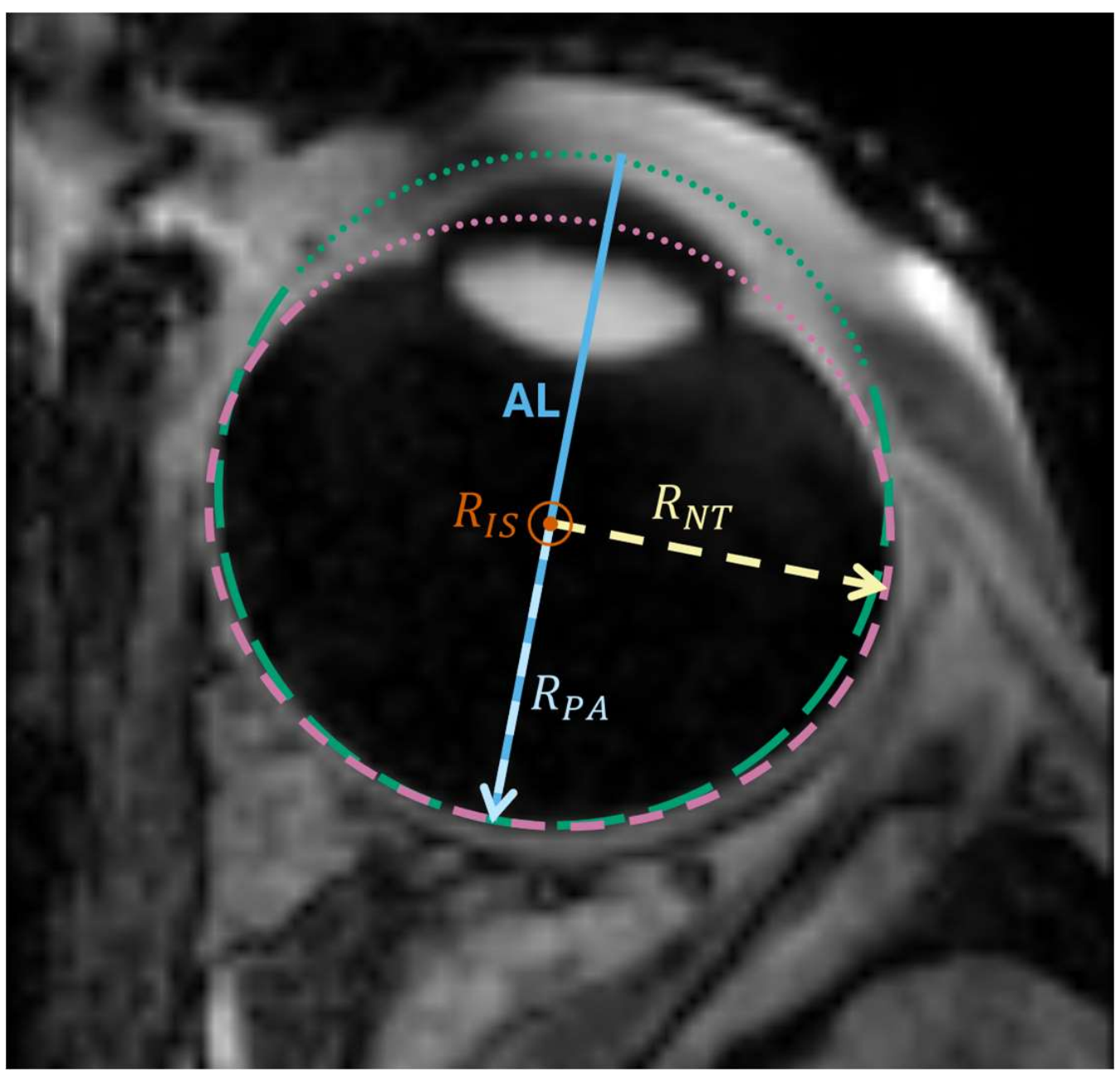

*Figure 1 Illustration of the included parameters shown in the inferior-superior plane of a T1-weighted MRI for a phakic eye. The green and pink circles correspond to the 12 mm spherical retinal curvature as reported in literature and the retinal curvature reported by van Vught et al. [29] , respectively. The dashed part shows 240 degrees of the retina used to fit the model. The sky blue continuous line represents the axial length (24.31 mm), defined as length from cornea to inner limiting membrane, measured with the Lenstar, which is in agreement with the MRI [29]. The yellow and light blue dashed lines provide the $R_{NT}$ (12.24 mm) and $R_{PA}$ (10.82 mm) radii in the nasal-temporal and posterior-anterior directions, respectively. The orange circle represents the location of the inferior-superior radius ($R_{IS}$=11.55 mm).*

*Models*

The large standard deviation in the retinal curvature reported in literature, suggests that a linear regression model would not sufficiently describe the distribution in the healthy population. Therefore, a more complex Gaussian Model was fitted on the data. Since it is unknown whether the relationship between axial length and retinal curvature necessitates two Gaussian components, both a Single Gaussian Model (SGM) [21] and a two-component MMGM [22] were evaluated. The SGM and MGMM are implemented in Python (v 3.13) using the scipy (v 1.15.2) and scikit-learn (v 1.6.1), respectively.

A conditional version of the retinal curvature model, based on a given axial length, was derived such that the retinal model can integrated into the existing SyntEyes model [22] (see Supplement 1 for the mathematical derivation). Using this formalism, 3D SyntEyes can be constructed by first generating the conventional SyntEyes parameters and then calculating the matching retinal radii for each eye, using the conditional retinal GM.

In this paper, both the non-conditional models, which do not take the axial length as input, and the conditional models, which take the axial length into consideration, were evaluated.

*Statistics*

Before fitting the models on the data, it was determined how many eyes in the measured dataset approximately followed the 12 mm sphere (Euclidian distance between sphere and ellipsoid $\leq$ 0.5mm), to underscore the importance of creating a more realistic model. Additionally, the effect of including pseudophakic eyes was investigated with a Bonferroni corrected Mann-Whitney-U test.

The performance of each model was tested with five-split cross-validation. Given the limited sample size, this approach was chosen to avoid overfitting and improve the generalizability of the resulting models [36]. Data splitting was performed randomly with the sole constraint of an equal distribution of pseudophakic eyes, resulting in four splits of 75-19 and one split 76-18 of training-validation data points. A Bonferroni corrected Mann-Whitney-U test was performed across the different splits to access whether the resulting training and validation sets formed similar distributions.

The trained models for the different splits were then used to generate datasets to compare the models with the measured data. For each eye included from the measured dataset in the split 100 eyes were generated with the corresponding model. This approach was chosen such that the generated dataset clearly represents the models' distribution.

Model performance was quantified with three measures. First, the difference in median was calculating by subtracting the model's median from the measured median. After that, the model's variance was compared to the measured variance using an f-test of equality of variance. Lastly, the overlap in Probability Density Functions (PDFs) between the data generated by the models and the training and validation data. A Gaussian kernel was used, with Scott's rule adopted to determine the appropriate bandwidth [37]. As the small dataset size limits the stability of the metrics, particularly for the validation datasets, the metrics are not determined for each split separately but on the entire generated cohort.

Lastly, the final model was assessed using the median absolute distance in mm between the radius Cumulative Density Functions (CDFs) of the measured and generated validation data, within the 0.05-0.95 cumulative range, excluding tail influences.

*Retinal model application*

The resulting average SGM model was integrated in the open-source ophthalmic raytracing platform Visisipy [38] to investigate the effect of the retinal model on the resulting relative peripheral refraction. The Navarro wide-angle eye [27] was taken as the basis and the relative refraction at 40° was calculated for 1000 conditional SGM eyes with the same axial length (23.92 mm) but varying retinal radii.

# Results

*Population cohort*

*Table 1 Characteristics of the dataset. In this table, IOL represents the number of participants with an intraocular lens implantation. The AL, $R_{NT}$, $R_{IS}$, and $R_{PA}$ represent the axial length and the retinal radii in the nasal-temporal, inferior-superior, and posterior-anterior direction, respectively.*

| Measured data | Median (min-max) |
|---|---|
| ***Participants*** | 94 |
| ***Sex*** (female) | 65 |
| ***IOL*** | 61 |
| ***Eye*** (left) | 65 |
| ***Age*** [years] | 63.2 (18.8-82.6) |
| ***AL*** [mm] | 23.6 (20.5-28.4) |
| $R_{NT}$ [mm] | 11.9 (10.4-14.6) |
| $R_{IS}$ [mm] | 11.6 (10.3-13.0) |
| $R_{PA}$ [mm] | 10.4 (9.2-12.4) |

The total cohort consisted of 94 individuals with a wide range in age (19–83 years) and axial length (20.5–28.4 mm). About two-thirds of the subjects were pseudophakic, and other descriptors of the cohort, such as the retinal radii, can be found in Table 1. In this dataset, only 3/94 (3%) of the retinal ellipsoids could be approximated as a sphere with a 12 mm radius. The Mann-Whitney-U test demonstrated no significant differences for any of the study parameters between the phakic and pseudophakic subjects ($p > 0.05/4 = 0.0125$, Bonferroni correction).

The distribution of axial length and the posterior-anterior radius resembles a two-component Gaussian distribution, while the nasal-temporal and inferior-superior radii more closely resemble a single-component Gaussian distribution (Figure 2). Although all four studied parameters are correlated, the strongest correlation was observed between axial length and the posterior-anterior radius ($r_{pearson} = 0.862$).

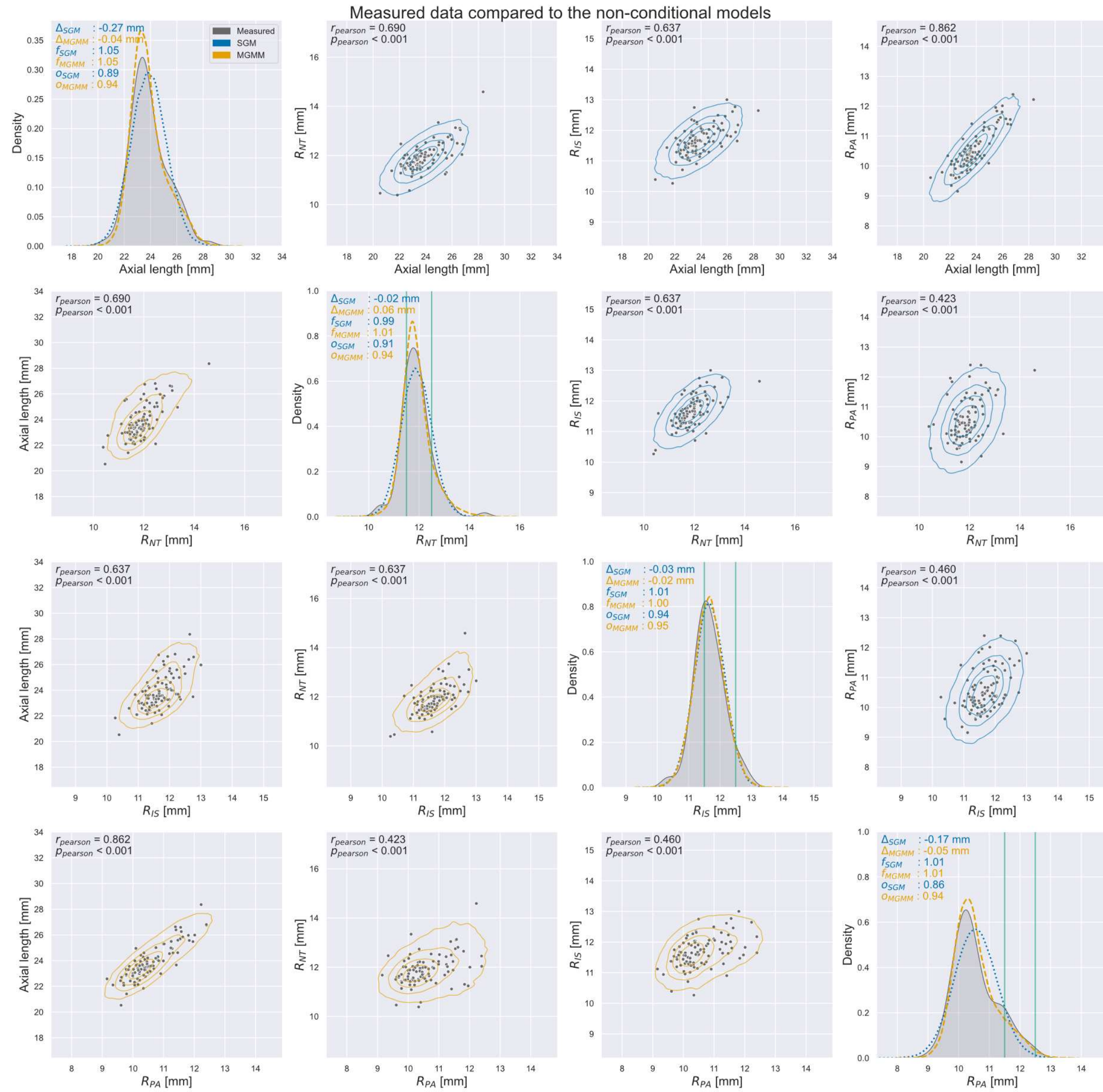

*Figure 2 The diagonal gives the distribution of the measured, the non-conditional SGM-generated, and MGMM-generated data for each biometric parameter in grey, blue, and orange, respectively. Furthermore, in the radii plots, green vertical lines are shown at 12 ± 0.5 mm representing, the neighborhood of the 12 mm radii from literature. Additionally, the median distance (Δ), f-test (f) and overlap (o) are shown. The other plots provide insight into the relationship between the different biometric parameters, with the grey dots representing the measured data. The blue and orange lines are contour plots of the SGM and MGMM across all 5 splits combined, respectively. The outermost contour includes 95% of the generated data, and the inner contours represent 75%, 50%, and 25%. In the non-diagonal images, the Pearson correlation and corresponding p-value are provided.*

The five training and validation datasets showed similar distributions as the Mann-Whitney-U test was not significant for all parameters and between every split ($p > 0.05/40 = 0.00125$, Bonferroni correction). Further details of the splits are provided in Supplement 3.

*Table 2 The overlap between the PDFs, the f-test of the measured data and eyes generated by the different models for the biometric parameters. This includes the axial length (AL) and the retinal radii in the nasal-temporal ($R_{NT}$), inferior-superior ($R_{IS}$), and posterior-anterior ($R_{PA}$) direction. There was no significant difference in the variance (p > 0.05) between the measured and generated data.*

| | | $\Delta_{median}$ [mm] | F-test (p-value) | Overlap [-] | $\Delta_{median}$ [mm] | F-test (p-value) | Overlap [-] |
|---|---|---|---|---|---|---|---|
| **Training** | **Parameters** | **SGM** | | | **MGMM** | | |
| | | **Non-conditional model** | | | | | |
| | ***AL*** | -0.27 | 1.05 (0.71) | 0.89 | -0.04 | 1.05 (0.73) | 0.94 |
| | $R_{NT}$ | -0.02 | 0.99 (0.99) | 0.91 | 0.06 | 1.01 (0.92) | 0.94 |
| | $R_{IS}$ | -0.03 | 1.01 (0.89) | 0.94 | -0.02 | 1.00 (0.96) | 0.95 |
| | $R_{PA}$ | -0.17 | 1.01 (0.93) | 0.86 | -0.05 | 1.01 (0.89) | 0.94 |
| | | **Conditional model** | | | | | |
| | $R_{NT}$ | 0.01 | 1.01 (0.90) | 0.93 | 0.05 | 1.00 (0.94) | 0.94 |
| | $R_{IS}$ | -0.01 | 1.01 (0.90) | 0.96 | -0.02 | 1.02 (0.84) | 0.95 |
| | $R_{PA}$ | -0.09 | 1.02 (0.85) | 0.93 | -0.04 | 1.01 (0.93) | 0.94 |
| **Validation** | | **Conditional model** | | | | | |
| | $R_{NT}$ | 0.01 | 1.02 (0.85) | 0.92 | - | - | - |
| | $R_{IS}$ | -0.01 | 1.01 (0.90) | 0.97 | - | - | - |
| | $R_{PA}$ | -0.10 | 1.02 (0.87) | 0.93 | - | - | - |

*Non-conditional models on the training dataset*

The non-conditional MGMM performs better for the axial length and posterior-anterior radii than the SGM as seen from both the median difference (AL -0.04 mm vs. -0.27 mm and $R_{PA}$ -0.05 mm vs. -0.17 mm) and overlap (AL 0.94 vs. 0.89 and $R_{PA}$ 0.94 vs. 0.86) while the f-test was similar for both models (Table 2). For the nasal-temporal and inferior-superior radii, the performance differences between the models was smaller and both models show a small median difference and overlap values over 91%. The slightly better performance of the MGMM was also confirmed with visual inspection (Figure 2). Furthermore, the distribution of the SGM stayed consistent across all splits, whereas the MGMM components differed between splits (Supplement 3), suggesting overfitting. If the splits would be assessed separately, similar conclusions would have been drawn (Supplement 4).

*Conditional models on the training dataset*

A comparable evaluation as on the non-conditional training data was performed on the conditional models, in which a set of retinal radii was determined for each measured axial length and compared with the measured radii. The SGM model performed clearly better than the non-conditional counterpart with median difference for the posterior-anterior reduced to -0.09 mm and overlap of 93%. For the other radii the median difference was still small (0.01 mm and -0.01 mm for $R_{NT}$ and $R_{IS}$, respectively) and the overlap had increased (0.93 and 0.96, respectively). For all radii, the f-test showed that there was no significant difference in variance between generated and measured data (p > 0.05). The performance of the MGMM stayed similar as the non-conditional counterpart. Furthermore, the performance between the SGM and MGMM was comparable which was also observed in the splits separately (Supplement 4).

*Conditional model on the validation dataset*

Given the similar performance of both models and the robustness of the less complex SGM, the validation was only performed on the SGM. Figure 3 provides the distributions, combined from all five splits, of the measured data and the conditional SGM model. The performance is comparable to the training data with median differences of 0.01, -0.01, and -0.10 mm for nasal-temporal, inferior-superior, and posterior-anterior, respectively. The f-test was still not significant ($p > 0.05$). Additionally, the overlap exceeded 92% for all three radii, suggesting that the models indeed capture the data correctly without overfitting. In the validation dataset, the performance for each split separately is unstable due to the small amount of datapoint in the dataset (Supplement 4). Furthermore, the median absolute distance between the CDF for the left-right, inferior-superior, and posterior-anterior direction are 0.06, 0.02, and 0.07 mm, respectively.

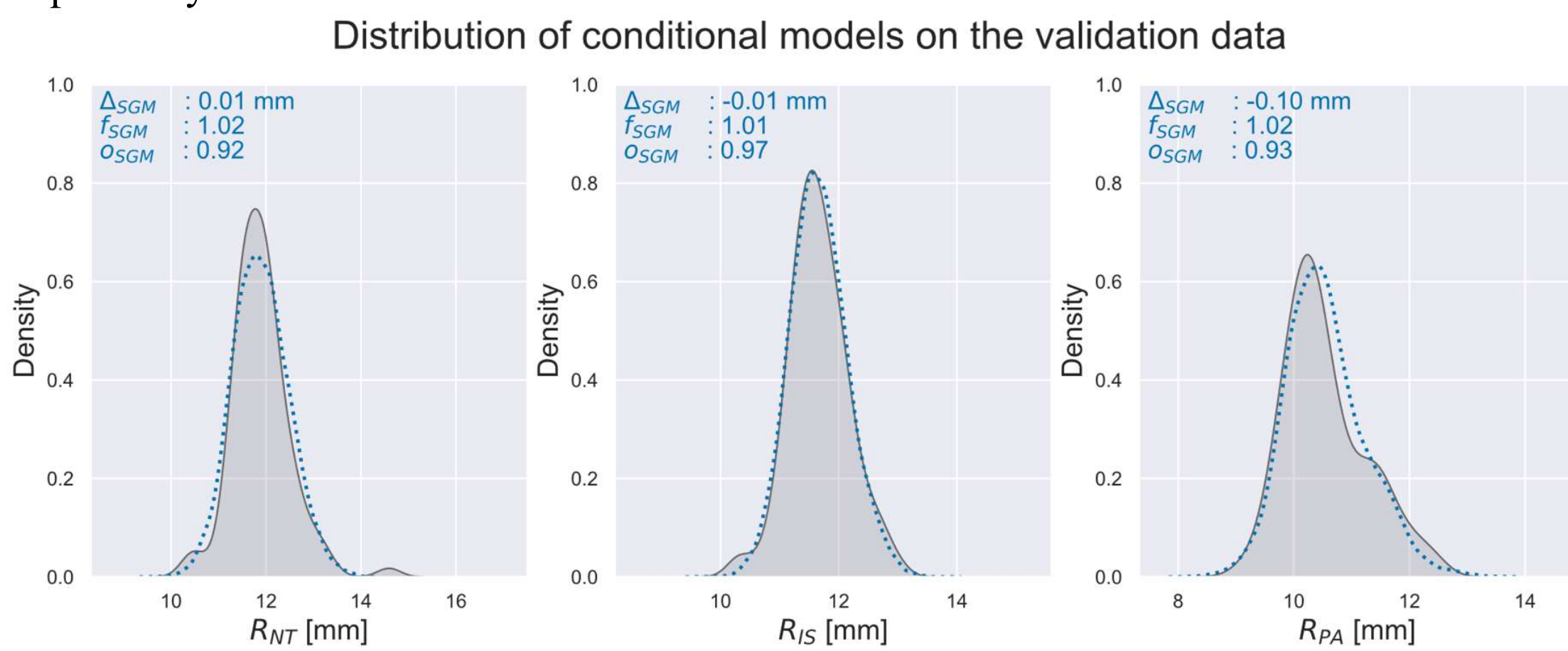

*Figure 3 Comparison between the distributions of $R_{NT}$, $R_{IS}$, and $R_{PA}$ for the conditional SGM model when the AL from the validation data was used as input. For the radii, the distributions of the measured validation data and the SGM are shown in gray and purple, respectively. In the corner, the median difference (Δ), f-test, and overlap provided.*

*Retinal model applied*

The 3D SyntEyes were successfully implemented in Visisipy and 1000 eyes were generated based on the Navarro schematic eye, but with retinal radii based on the average conditional SGM (Supplement 2). For its axial length of 23.92 mm the resulting standard deviation in retinal radii were 0.43, 0.38, and 0.35 in the nasal-temporal, inferior-superior, and posterior-anterior direction, respectively. When for these 1000 eyes the wavefront aberrations at 40° in the inferior-superior and posterior-anterior plane was calculated, this resulted in a median peripheral refraction of 0.99 D, with an inter-quartile range of 1.38 D (see Figure 4). The 12 mm spherical retinal curvature resulted in a peripheral refraction of 1.46 D.

# Discussion

In this study, a stochastic model was established for the relationship between axial length and three-dimensional retinal radii in healthy eyes. Both an SGM and an MGMM were evaluated as non-conditional and conditional (on the axial length) stochastic models. Consistent with previous work, the axial length distribution in our cohort exhibited a bi-Gaussian character, reflecting the presence of a myopic subgroup [39]. Despite this, a single Gaussian model proved sufficient to derive statistically representative retinal radii when conditioned on axial length. When the conditional SGM is combined with the previously published SyntEyes model [21,22], this approach enables the generation of three-dimensional eye models that more faithfully represent anatomical variation in the posterior segment of the healthy population.

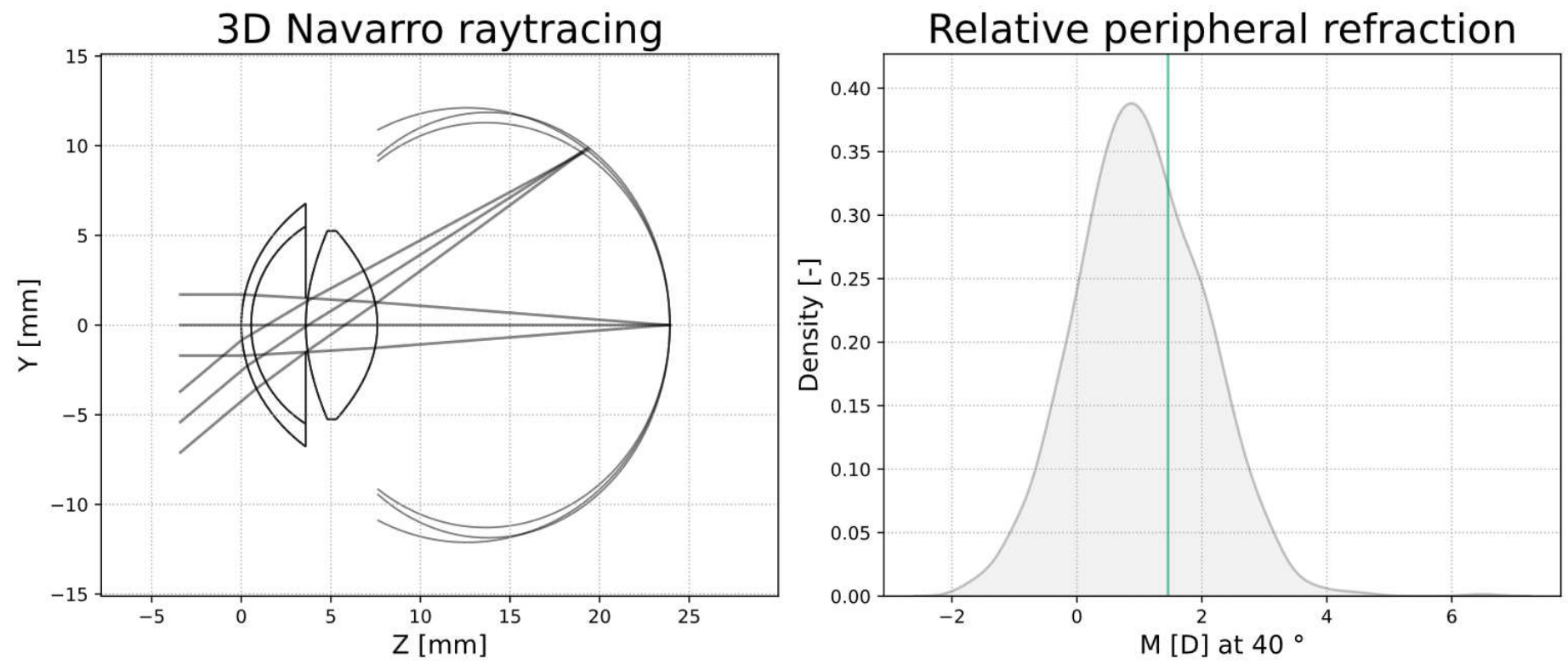

*Figure 4 Raytracing at 40° was performed on 1000 eyes based on Navarro wide-angle model with the conditional SGM. On the left, three different retina curvatures are shown which shows that the ray is affected by the shape at the peripheral retina. On the right, the PDF of the relative peripheral refraction for the 1000 eyes is provided with the green vertical line representing the results for the 12 mm spherical retinal curvature used in literature.*

The proposed conditional model demonstrated a close agreement with the measured data. Across all three retinal radii of curvature, the overlap between generated and measured distributions exceeded 92%, indicating that the model captures not only the average eye but also the clinically relevant spread of retinal geometries. Importantly, the absolute median CDF distance for each radius of curvature was at most 0.07 mm. When this value was compared to the standard deviation observed with the Navarro eye (minimal 0.35 mm), the CDF distance is small. Furthermore, the ray-tracing simulations showed that for an average eye the variation in retinal radii alone, results in a IQR of 1.38 D in peripheral refraction at 40° supporting the clinical relevance of a stochastic retinal model.

A limitation of this study is the relatively small dataset of 94 subjects, which precluded the use of an independent validation cohort and formal goodness-of-fit testing with metrics such as the two one-sided t-test employed in earlier SyntEyes studies [21,22]. Although the dataset spans a wide range of axial lengths and retinal curvatures, a larger sample size would allow independent validation and the exploration of more flexible statistical representations, including the t-distribution [40] the skewed-normal distribution [41], or non-parametric mixture components [42]. Such models may better capture the underlying distribution, particularly for myopic eyes, where axial length and, by extension, retinal curvature is expected to be more complex than a single Gaussian. Nevertheless, the strong performance of the present conditional Gaussian model suggests that it already captures the measured variation in the data without evidence of clinically relevant overfitting and so outperforms the 12 mm spherical retina often assumed. Furthermore, the cross-validation did allow to investigate the robustness of the model despite the lack of independency and showed that the MGMM model was less robust. This observation can be explained by the large increase in fitting parameters for the MGMM versus SGM for which more data is needed to create a stable result.

It is important to note that the proposed stochastic retinal model incorporates only axial length as an input. Although other biometric parameters, such as anterior chamber depth, corneal curvature, refractive error, and off-axis refraction, may also be associated with retinal curvature. These relationships are expected to be weaker than that of the axial length. This is further supported by the strong agreement between the generated and measured retinal radii.

The proposed three-dimensional statistical eye model enables several clinically and technologically relevant applications. In refractive and cataract surgery, the extended SyntEyes framework allows population-based optical simulations that incorporate not only variability in corneal topography and axial length but also the associated distribution of retinal curvatures. This is particularly relevant for the design and evaluation of intraocular lenses aimed at preserving or optimizing peripheral vision [43], for understanding perceived visual phenomena such as (negative) dysphotopsia [8,9], and for estimating the error in optical imaging modalities designed with the 12 mm assumption [34]. By accounting for realistic retinal geometry, the present model enables more comprehensive virtual cohorts for optical design studies, reducing reliance on simplified assumptions such as a fixed 12 mm retinal radius. To facilitate broader adoption in visual optics research, an online demo is created at demo.mreye.nl/synteyes, where 3D SyntEyes can be created with a retinal curvature determined using the axial length from the SyntEyes variables. The underlying code is available at https://github.com/MREYE-LUMC/3dsynteyes. These 3D SyntEyes have also been integrated into Visisipy, an open source raytracing platform specifically developed for visual optics simulations [38].

Beyond refractive surgery, the model supports applications in biomechanical modeling and ocular radiotherapy. The availability of realistic, population based posterior segment geometry provides a physiologically meaningful starting point for simulations of ocular growth and mechanical remodeling, for example in conditions such as keratoconus [25,44], myopia progression [31,33] and understanding the principles of crystalline lens wobbling [45]. In radiotherapy, statistical eye models may aid the development of more data-driven safety margins [46], particularly where uncertainties arise from mapping two-dimensional fundus images onto three-dimensional anatomical models used for treatment planning [34]. Finally, the availability of scalable, population-based 3D eye models facilitates generating large, anatomically plausible synthetic datasets for training AI algorithms in visual optics and ophthalmic imaging [47-49].

# Conclusions

In conclusion, the retinal curvature can be modeled accurately with a SGM conditioned on the axial length. This opens the path for more representative eye simulations across the peripheral retina.

# Supplement 1

The Gaussian Mixture models can be made conditional by determining a new mean, covariance, and weights for the unknown parameters with the following equations

$$\boldsymbol{\mu}_{u,k|\boldsymbol{x_m}} = \boldsymbol{\mu}_{u,k} + \boldsymbol{\Sigma}_{um,k}\boldsymbol{\Sigma}_{mm,k}^{-1}\left(\boldsymbol{x}_m - \boldsymbol{\mu}_{m,k}\right), \tag{1}$$

$$\boldsymbol{\Sigma}_{uu,k|\boldsymbol{x_m}} = \boldsymbol{\Sigma}_{uu,k} - \boldsymbol{\Sigma}_{um,k}\boldsymbol{\Sigma}_{mm,k}^{-1}\boldsymbol{\Sigma}_{um,k}^{T}\,, \tag{2}$$

and

$$\omega_{k|\boldsymbol{x_m}} = \frac{\omega_k \cdot \phi\left(\boldsymbol{x}_m|\boldsymbol{\mu}_{m,k}, \boldsymbol{\Sigma}_{mm,k}\right)}{\sum_l \omega_l \cdot \phi\left(\boldsymbol{x}_m|\boldsymbol{\mu}_{m,l}, \boldsymbol{\Sigma}_{mm,l}\right)} \tag{3}$$

[50-52]. In these equations, $\boldsymbol{\mu}$, $\boldsymbol{\Sigma}$, $\omega$, and $\boldsymbol{x}$ represent the mean, covariance, weights, and the known value, respectively. The $u$ and $m$ subscripts stand for the unknow and measured data while $k$ and $l$ stand of the k[th] and l[th] Gaussian component. Note that, for the Single Gaussian Model (SGM) there is only one Gaussian component and the weight will therefore be 1. Lastly, $\phi$ is a Gaussian Probability Density Function (PDF) with as mean and covariance the parameters after the bar (|) and this is measured at the location of the parameter before the bar.

These equations can be rewritten for the 3D-retinal curvature SGM where the axial length is measured and the radii are the unknown variables. In this case, the measured value is the axial length (AL) with the mean ($\mu_{AL}$) and the variance ($\sigma_{AL}^2$).

The unknown variables are the radii $\begin{bmatrix} R_{NT} \\ R_{IS} \\ R_{PA} \end{bmatrix}$, with the means $\mu_{R_i}$ (with i=NT, IS, PA depending on the radii of interest). The covariances are $\sigma_{R_i,AL}$ and $\sigma_{R_i,R_j}$ between the i[th] radii and the axial length, and the i[th] radii and the j[th] radii, respectively.

Resulting in the following conditional distribution parameters:

$$\begin{bmatrix} \mu_{R_{NT}|AL} \\ \mu_{R_{IS}|AL} \\ \mu_{R_{PA}|AL} \end{bmatrix} = \begin{bmatrix} \mu_{R_{NT}} \\ \mu_{R_{IS}} \\ \mu_{R_{PA}} \end{bmatrix} + \frac{(AL - \mu_{AL})}{\sigma_{AL}^2} \begin{bmatrix} \sigma_{R_{NT},AL} \\ \sigma_{R_{IS},AL} \\ \sigma_{R_{PA},AL} \end{bmatrix}, \tag{4}$$

$$\Sigma_{R|AL} = \begin{bmatrix} \sigma_{R_{NT}|AL}^2 & \sigma_{R_{NT},R_{IS}|AL} & \sigma_{R_{NT},R_{PA}|AL} \\ \sigma_{R_{NT},R_{IS}|AL} & \sigma_{R_{IS}|AL}^2 & \sigma_{R_{IS},R_{PA}|AL} \\ \sigma_{R_{NT},R_{PA}|AL} & \sigma_{R_{IS},R_{PA}|AL} & \sigma_{R_{PA}|AL}^2 \end{bmatrix}, \tag{5}$$

where

$$\sigma_{R_i|AL}^2 = \sigma_{R_i}^2 - \frac{\sigma_{R_i,AL}^2}{\sigma_{AL}^2}, \tag{6}$$

$$\sigma_{R_i,R_j|AL} = \sigma_{R_i,R_j} - \frac{\sigma_{R_i,AL}\sigma_{R_j,AL}}{\sigma_{AL}^2}\,. \tag{7}$$

## Supplement 2

The final mean matrix $\boldsymbol{\mu}$ and covariance matrix $\boldsymbol{\Sigma}$, which is calculated by taking the average of the five models created for each split, are provided in Table S2.1 and Table S2.2, respectively. These values are needed to generate a cohort of realistic retinal curvatures. If this is directly implemented, a non-conditional SGM is implemented. On the other hand, if these values are implemented in combination with Supplement 1 the conditional SGM can be designed (also see demo.mreye.nl/synteyes).

*Table S2.1 Mean matrix ($\boldsymbol{\mu}$) for the Single Gaussian Model. The AL, $R_{NT}$, $R_{IS}$, and $R_{PA}$ represent the axial length and the retinal radii in the nasal-temporal, inferior-superior, and posterior-anterior direction, respectively.*

| SGM | AL | $R_{NT}$ | $R_{IS}$ | $R_{PA}$ |
|---|---|---|---|---|
| $\boldsymbol{\mu}$ [mm] | 23.85 | 11.89 | 11.66 | 10.55 |

*Table S2.2 Covariance matrix ($\boldsymbol{\Sigma}$) for the Single Gaussian Model. The AL, $R_{NT}$, $R_{IS}$, and $R_{PA}$ represent the axial length and the retinal radii in the nasal-temporal, inferior-superior, and posterior-anterior direction, respectively.*

| SGM $\boldsymbol{\Sigma}$ | AL | $R_{NT}$ | $R_{IS}$ | $R_{PA}$ |
|---|---|---|---|---|
| AL | 1.82 | 0.56 | 0.42 | 0.81 |
| $R_{NT}$ | 0.56 | 0.36 | 0.19 | 0.18 |
| $R_{IS}$ | 0.42 | 0.19 | 0.24 | 0.16 |
| $R_{PA}$ | 0.81 | 0.18 | 0.16 | 0.48 |

# Supplement 3

The axial length and radii of the retinal curvature are compared over the different splits to investigate if these parameters are similar over the splits. After that, the resulting Single Gaussian Models and Multi Gaussian Mixture Models for the individual splits are provided.

*Table S3.1 For each split in the training dataset, the median and interquartile range are provided. The AL, $R_{NT}$, $R_{IS}$, and $R_{PA}$ represent the axial length and the retinal radii in the nasal-temporal, inferior-superior, and posterior-anterior direction, respectively. It can be seen that these values are very similar over the different splits. This was also verified with the Mann-Whitney U test which did not show a significant difference between the splits.*

| Training dataset | Split 1 | Split 2 | Split 3 | Split 4 | Split 5 |
|---|---|---|---|---|---|
| *Split size* | 75 | 75 | 75 | 75 | 76 |
| ***AL [mm]*** | | | | | |
| Median [mm] | 23.56 | 23.56 | 23.57 | 23.56 | 23.62 |
| Inter Quartile Range [mm] | 1.62 | 1.34 | 1.58 | 1.43 | 1.89 |
| $R_{NT}$ ***[mm]*** | | | | | |
| Median [mm] | 11.88 | 11.84 | 11.88 | 11.87 | 11.89 |
| Inter Quartile Range [mm] | 0.63 | 0.61 | 0.61 | 0.62 | 0.72 |
| $R_{IS}$ ***[mm]*** | | | | | |
| Median [mm] | 11.63 | 11.60 | 11.63 | 11.65 | 11.65 |
| Inter Quartile Range [mm] | 0.64 | 0.63 | 0.67 | 0.64 | 0.58 |
| $R_{PA}$ ***[mm]*** | | | | | |
| Median [mm] | 10.38 | 10.38 | 10.36 | 10.38 | 10.38 |
| Inter Quartile Range [mm] | 0.93 | 0.74 | 0.87 | 0.79 | 1.14 |

*Table S3.2 For each split in the validation dataset, the median and interquartile range are provided. The AL, $R_{NT}$, $R_{IS}$, and $R_{PA}$ represent the axial length and the retinal radii in the nasal-temporal, inferior-superior, and posterior-anterior direction, respectively. It can be seen that these values are still similar over the different splits. Although, that the differences are larger due to the small split size, still the Mann-Whitney U test did not show a significant difference between the splits.*

| Validation dataset | Split 1 | Split 2 | Split 3 | Split 4 | Split 5 |
|---|---|---|---|---|---|
| *Split size* | 19 | 19 | 19 | 19 | 18 |
| ***AL [mm]*** | | | | | |
| Median [mm] | 23.59 | 23.70 | 23.59 | 23.66 | 23.31 |
| Inter Quartile Range [mm] | 1.00 | 2.06 | 1.78 | 1.79 | 1.14 |
| $R_{NT}$ ***[mm]*** | | | | | |
| Median [mm] | 11.80 | 12.00 | 11.84 | 11.88 | 11.84 |
| Inter Quartile Range [mm] | 0.64 | 0.64 | 0.65 | 0.64 | 0.43 |
| $R_{IS}$ ***[mm]*** | | | | | |
| Median [mm] | 11.64 | 11.75 | 11.65 | 11.60 | 11.43 |
| Inter Quartile Range [mm] | 0.52 | 0.71 | 0.41 | 0.65 | 0.68 |
| $R_{PA}$ ***[mm]*** | | | | | |
| Median [mm] | 10.36 | 10.56 | 10.38 | 10.35 | 10.38 |
| Inter Quartile Range [mm] | 0.48 | 1.14 | 1.16 | 1.18 | 0.56 |

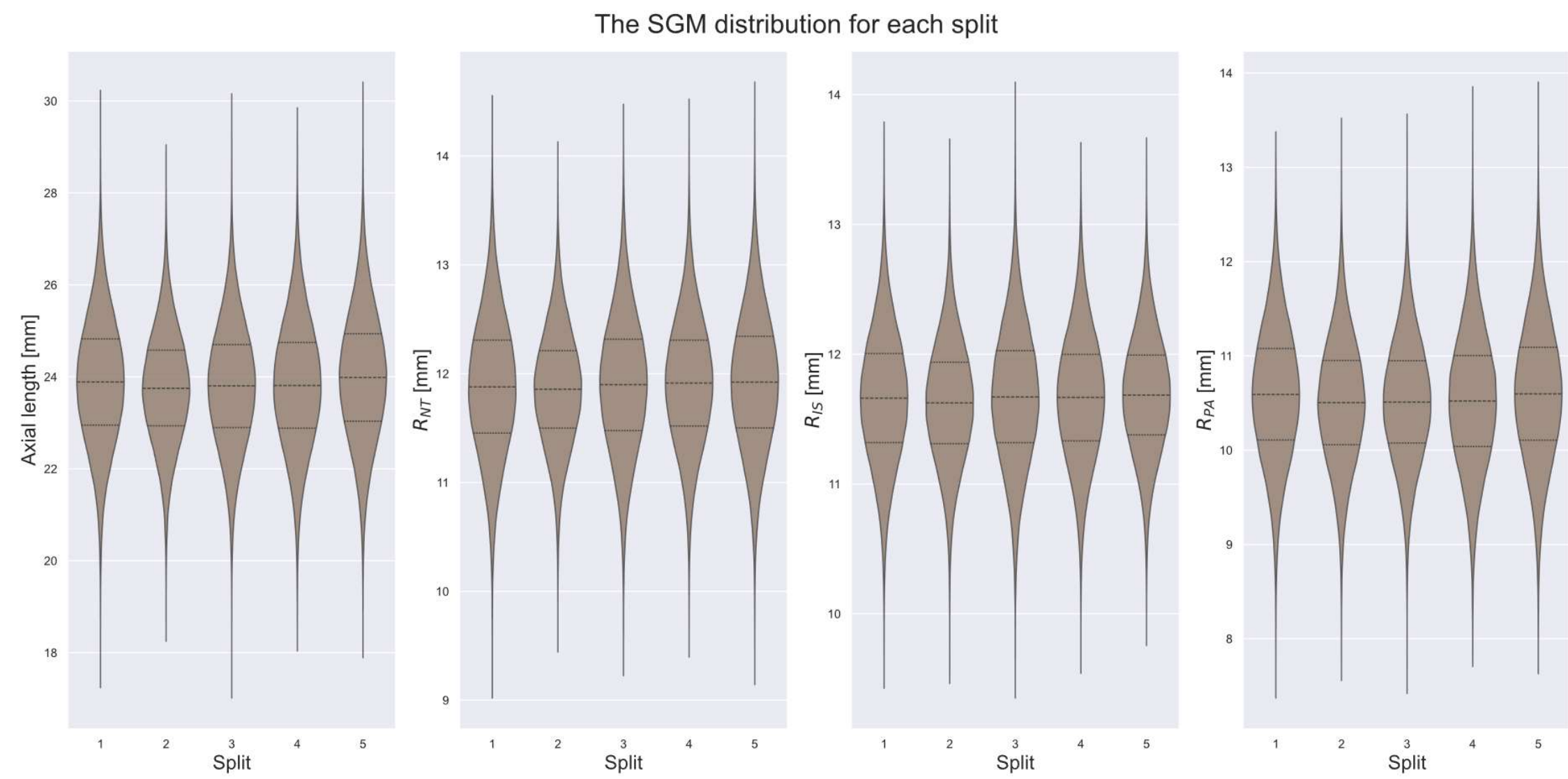

*Figure S3.1 Violin plots of the distribution of the non-conditional SGM for each split. The SGM stays stable over the different splits with only minor deviations in median and interquartile range over the different splits.*

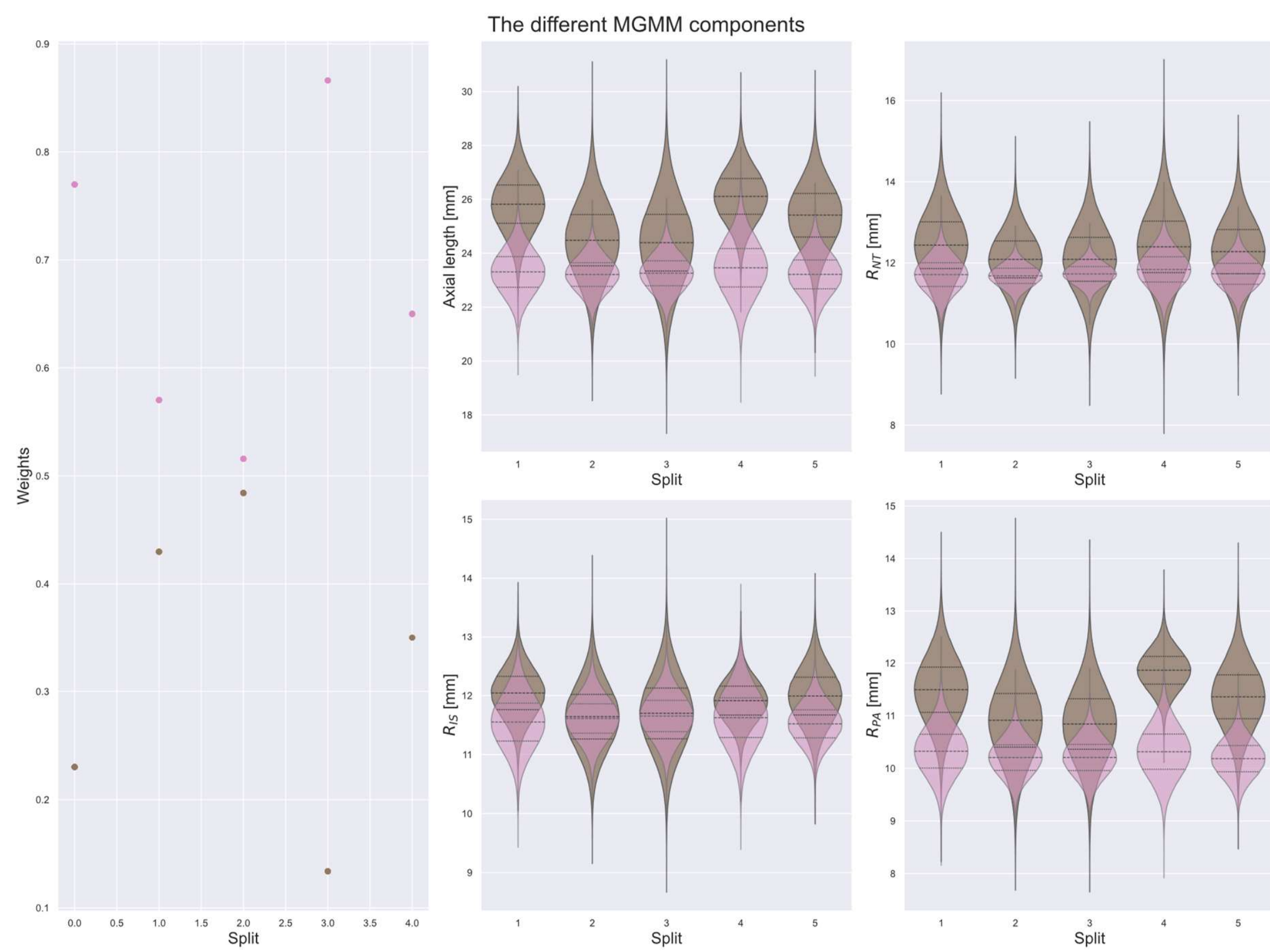

*Figure S3.2 Violin plots of the components of the non-conditional MGMM for each split. For each split, all the pink violins represent the same component and the brown violins the other component. The pink components stays stable over the different splits while the brown component depends on the chosen split. Therefore, the brown component might be modelling noise instead of an underlying component and might be overfitting the data.*

# Supplement 4

The performance for each model separately over each split for the investigated parameters is provided. Each table consists of a different combination of the models with either the training or the validation dataset.

*Table S4.1 For each split in training data, the performance of the non-conditional Single Gaussian Model (SGM) was evaluated with the median difference ($\Delta_{median}$), f-test, and overlap. The AL, $R_{NT}$, $R_{IS}$, and $R_{PA}$ represent the axial length and the retinal radii in the nasal-temporal, inferior-superior, and posterior-anterior direction, respectively. Over the different splits, the metrices show some variation. However, the model still captured the $R_{NT}$ and $R_{IS}$ distribution accurately, while it had more difficulties to capture the AL and $R_{PA}$.*

| Non-conditional SGM training | Split 1 | Split 2 | Split 3 | Split 4 | Split 5 |
|---|---|---|---|---|---|
| ***AL [mm]*** | | | | | |
| $\Delta_{median}$ [mm] | -0.27 | -0.21 | -0.23 | -0.30 | -0.39 |
| F-test (p-value) | 1.03 (0.80) | 1.06 (0.67) | 1.01 (0.91) | 1.07 (0.66) | 1.05 (0.74) |
| Overlap | 0.89 | 0.88 | 0.91 | 0.89 | 0.89 |
| $R_{NT}$ ***[mm]*** | | | | | |
| $\Delta_{median}$ [mm] | 0.03 | -0.02 | -0.01 | -0.05 | -0.05 |
| F-test (p-value) | 0.99 (0.98) | 1.01 (0.91) | 0.98 (0.96) | 1.01 (0.91) | 1.00 (0.96) |
| Overlap | 0.90 | 0.92 | 0.90 | 0.90 | 0.90 |
| $R_{IS}$ ***[mm]*** | | | | | |
| $\Delta_{median}$ [mm] | -0.02 | -0.03 | -0.05 | -0.03 | -0.04 |
| F-test (p-value) | 1.02 (0.88) | 1.05 (0.74) | 1.00 (0.95) | 1.00 (0.94) | 1.02 (0.88) |
| Overlap | 0.94 | 0.94 | 0.94 | 0.94 | 0.94 |
| $R_{PA}$ ***[mm]*** | | | | | |
| $\Delta_{median}$ [mm] | -0.19 | -0.15 | -0.15 | -0.16 | -0.22 |
| F-test (p-value) | 1.01 (0.92) | 1.02 (0.85) | 1.00 (0.95) | 1.00 (0.94) | 1.01 (0.93) |
| Overlap | 0.86 | 0.85 | 0.88 | 0.87 | 0.85 |

*Table S4.2 For each split in training data, the performance of the non-conditional Multi Gaussian Mixture Model (MGMM) was evaluated with the median difference ($\Delta_{median}$), f-test, and overlap. The AL, $R_{NT}$, $R_{IS}$, and $R_{PA}$ represent the axial length and the retinal radii in the nasal-temporal, inferior-superior, and posterior-anterior direction, respectively. Over the different splits, the metrices show some variation. However, the model captured all parameters accurately with overlap values exceeding 0.9.*

| Non-conditional MGMM training | Split 1 | Split 2 | Split 3 | Split 4 | Split 5 |
|---|---|---|---|---|---|
| ***AL [mm]*** | | | | | |
| $\Delta_{median}$ [mm] | -0.06 | 0.01 | >-0.01 | -0.11 | -0.10 |
| F-test (p-value) | 1.04 (0.77) | 1.08 (0.59) | 1.02 (0.87) | 1.07 (0.66) | 1.01 (0.90) |
| Overlap | 0.94 | 0.92 | 0.93 | 0.96 | 0.94 |
| $R_{NT}$ **[mm]** | | | | | |
| $\Delta_{median}$ [mm] | 0.07 | 0.07 | 0.08 | -0.02 | 0.04 |
| F-test (p-value) | 1.04 (0.77) | 1.02 (0.87) | 1.00 (0.94) | 1.03 (0.82) | 0.97 (0.89) |
| Overlap | 0.94 | 0.90 | 0.91 | 0.94 | 0.94 |
| $R_{IS}$ **[mm]** | | | | | |
| $\Delta_{median}$ [mm] | -0.02 | -0.03 | -0.05 | -0.03 | -0.01 |
| F-test (p-value) | 0.99 (0.97) | 1.03 (0.82) | 0.99 (0.99) | 1.02 (0.85) | 1.00 (0.94) |
| Overlap | 0.94 | 0.94 | 0.94 | 0.93 | 0.97 |
| $R_{PA}$ **[mm]** | | | | | |
| $\Delta_{median}$ [mm] | -0.10 | -0.01 | -0.06 | -0.03 | -0.05 |
| F-test (p-value) | 1.01 (0.92) | 1.07 (0.62) | 1.02 (0.86) | 1.00 (0.96) | 0.99 (0.99) |
| Overlap | 0.92 | 0.94 | 0.94 | 0.95 | 0.94 |

*Table S4.3 For each split in training data, the performance of the conditional Single Gaussian Model (SGM) was evaluated with the median difference ($\Delta_{median}$), f-test, and overlap. The $R_{NT}$, $R_{IS}$, and $R_{PA}$ represent the retinal radii in the nasal-temporal, inferior-superior, and posterior-anterior direction, respectively. Over the different splits, the metrices show some variation. However, the model captured the distributions of each split better than the non-conditional counterpart and all overlap values now exceed 0.9.*

| Conditional SGM training | Split 1 | Split 2 | Split 3 | Split 4 | Split 5 |
|---|---|---|---|---|---|
| $R_{NT}$ **[mm]** | | | | | |
| $\Delta_{median}$ [mm] | 0.04 | 0.01 | -0.01 | -0.01 | 0.01 |
| F-test (p-value) | 1.02 (0.88) | 0.99 (1.00) | 1.02 (0.86) | 1.02 (0.86) | 1.02 (0.86) |
| Overlap | 0.92 | 0.95 | 0.91 | 0.93 | 0.93 |
| $R_{IS}$ **[mm]** | | | | | |
| $\Delta_{median}$ [mm] | -0.03 | -0.01 | -0.03 | -0.01 | -0.02 |
| F-test (p-value) | 1.02 (0.86) | 1.01 (0.89) | 1.02 (0.88) | 1.00 (0.95) | 1.02 (0.85) |
| Overlap | 0.95 | 0.95 | 0.95 | 0.95 | 0.97 |
| $R_{PA}$ **[mm]** | | | | | |
| $\Delta_{median}$ [mm] | -0.13 | -0.06 | -0.11 | -0.06 | -0.13 |
| F-test (p-value) | 1.02 (0.88) | 1.01 (0.89) | 1.02 (0.88) | 1.03 (0.80) | 1.03 (0.80) |
| Overlap | 0.92 | 0.95 | 0.92 | 0.93 | 0.91 |

*Table S4.4 For each split in training data, the performance of the conditional Multi Gaussian Mixture Model (MGMM) was evaluated with the median difference ($\Delta_{median}$), f-test, and overlap. The $R_{NT}$, $R_{IS}$, and $R_{PA}$ represent the retinal radii in the nasal-temporal, inferior-superior, and posterior-anterior direction, respectively. Over the different splits, the metrices show some variation. However, the model captured all parameters accurately with overlap values exceeding 0.9. The conditional MGMM has similar performance as the non-conditional counterpart.*

| Conditional MGMM training | Split 1 | Split 2 | Split 3 | Split 4 | Split 5 |
|---|---|---|---|---|---|
| **$R_{NT}$ [mm]** | | | | | |
| $\Delta_{median}$ [mm] | 0.07 | 0.06 | 0.07 | -0.01 | 0.05 |
| F-test (p-value) | 1.00 (0.94) | 0.98 (0.97) | 1.00 (0.94) | 1.02 (0.88) | 1.02 (0.88) |
| Overlap | 0.94 | 0.91 | 0.90 | 0.95 | 0.94 |
| **$R_{IS}$ [mm]** | | | | | |
| $\Delta_{median}$ [mm] | -0.04 | -0.03 | -0.05 | -0.02 | <0.01 |
| F-test (p-value) | 1.01 (0.89) | 1.04 (0.75) | 0.99 (0.98) | 1.06 (0.67) | 1.02 (0.86) |
| Overlap | 0.94 | 0.95 | 0.93 | 0.93 | 0.97 |
| **$R_{PA}$ [mm]** | | | | | |
| $\Delta_{median}$ [mm] | -0.09 | -0.02 | -0.05 | -0.01 | -0.04 |
| F-test (p-value) | 1.03 (0.84) | 1.01 (0.89) | 1.03 (0.83) | 0.98 (0.96) | 0.99 (0.99) |
| Overlap | 0.92 | 0.95 | 0.94 | 0.92 | 0.94 |

*Table S4.5 For each split in validation data, the performance of the conditional Single Gaussian Model (SGM) was evaluated with the median difference ($\Delta_{median}$), f-test, and overlap. The $R_{NT}$, $R_{IS}$, and $R_{PA}$ represent the retinal radii in the nasal-temporal, inferior-superior, and posterior-anterior direction, respectively. Over the different splits, the metrices show large variation. These variations are caused by the small validation datasets. Therefore, these metrices can not be used for these small datasets.*

| Conditional SGM validation | Split 1 | Split 2 | Split 3 | Split 4 | Split 5 |
|---|---|---|---|---|---|
| **$R_{NT}$ [mm]** | | | | | |
| $\Delta_{median}$ [mm] | <0.01 | 0.04 | -0.09 | -0.07 | 0.15 |
| F-test (p-value) | 0.52 (0.10) | 1.79 (0.04) | 0.71 (0.40) | 1.45 (0.20) | 0.87 (0.78) |
| Overlap | 0.84 | 0.91 | 0.91 | 0.81 | 0.86 |
| **$R_{IS}$ [mm]** | | | | | |
| $\Delta_{median}$ [mm] | 0.06 | 0.03 | -0.05 | -0.10 | -0.11 |
| F-test (p-value) | 0.76 (0.50) | 1.19 (0.53) | 0.35 (0.01) | 1.12 (0.64) | 2.64 (<0.01) |
| Overlap | 0.94 | 0.88 | 0.80 | 0.90 | 0.73 |
| **$R_{PA}$ [mm]** | | | | | |
| $\Delta_{median}$ [mm] | -0.13 | -0.09 | -0.12 | -0.13 | 0.11 |
| F-test (p-value) | 0.92 (0.89) | 0.86 (0.75) | 1.67 (0.08) | 0.95 (0.97) | 0.88 (0.81) |
| Overlap | 0.88 | 0.89 | 0.82 | 0.93 | 0.93 |